# La evaluación de la investigación: España suspende

Alonso Rodríguez Navarro

*Departamento de Biotecnología-Biología Vegetal, Universidad Politécnica de Madrid; Departamento de Estructura de la Materia, Física Térmica y Electrónica, Universidad Complutense de Madrid.*

**1. Introducción**

Desde hace décadas, gobiernos, economistas y empresarios, entre otros, repiten constantemente la importancia que tendría para España cambiar el modelo productivo, incrementando nuestra actividad industrial y tecnológica. Sin embargo, este atractivo proyecto es notablemente complejo. Basta ojear cualquier estudio sobre el tema, por ejemplo el del Círculo de Empresarios[1], para apreciar esa complejidad. Y si se analiza cuidadosamente cómo los sectores competitivos en España han llegado a su buen nivel, por ejemplo, la industria auxiliar del automóvil o la industria cerámica, la conclusión es similar.

La idea anterior conecta con una más global de la economía del conocimiento[2] que nos lleva directamente a situar a la actividad investigadora en el centro del proyecto. Sería ingenuo pensar que la investigación por sí misma puede cambiar el modelo productivo de un país, pero es razonable pensar que sin una investigación competitiva España va a tener muy difícil cambiar el modelo productivo. Es una condición necesaria aunque no suficiente. Por ejemplo, en el momento de escribir este artículo, el gobierno hace propaganda sobre la instalación de una planta de baterías eléctricas para la automoción, ignorando nuestro bajo nivel investigador en un área muy dinámica y competitiva como es la de ese sector. Esto nos lleva a una estricta dependencia tecnológica y a que España solo ponga la mano de obra y el solar para los edificios. Esto es mejor que nada, pero conlleva riesgos.

Al margen de la importancia que pueda tener la investigación para cambiar un sistema productivo, hay que destacar que mientras lo segundo es complejo y dependiente de numerosos factores, alcanzar un buen nivel del sistema público de investigación es mucho más sencillo. Mejorar la investigación en el sector privado es complicado por el reducido tamaño de la mayor parte de las empresas de nuestro tejido industrial, pero para mejorarla en el sector público solo hace falta voluntad política y saber hacerlo. Precisamente por la pequeña dimensión de muchas empresas en el sector privado, la importancia de la mejora de la investigación en el sector público se acrecienta, ya que este sector tiene que actuar como base, como referencia y como catalizador del sector privado.

En resumen, España necesita una investigación pública competitiva y conseguir esto solo depende de la voluntad política y del acierto de quienes se ocupan de ello, como ya se ha dicho, pero teniendo en cuenta que inversión no es sinónimo de competitividad. El tamaño del sistema sí depende de la inversión y, por ello, la dimensión del sistema entra dentro de la discusión política. Por el contrario, la calidad del sistema no debería ser objeto de discusión. Se podrá discutir si se quiere un sistema con 2.000 o 7.000 investigadores por millón de habitantes, pero no la eficiencia del sistema. Un sistema de investigación que no produce avances del conocimiento no es útil, cualquiera que sea su dimensión.

Un origen frecuente de discusión al hablar de un sistema de investigación es mezclarlo con el de innovación. Las innovaciones pueden ser radicales, basadas en importantes avances del conocimiento que dan origen a un nuevo producto o a un producto muy mejorado, o incrementales, basadas en pequeñas mejoras realizadas en un proceso o sobre un producto[3]. Las innovaciones radicales dependen directamente de la investigación, pública o privada, pero la dependencia de las innovaciones incrementales de la investigación pública es mucho menor.

Llevar una innovación incremental al mercado es un proceso complejo que depende directamente de las empresas y de su investigación; el papel de la investigación pública en este proceso no es evidente ya que puede acabar siendo una financiación encubierta que altera las reglas de la competencia. Comparando la magnitud del sector productivo y del sector público de investigación, la actuación del segundo en las innovaciones incrementales, si no está muy bien pensada, solo alcanzará a unas pocas empresas, en perjuicio de las demás.

Por otra parte, analizar la competitividad del sistema español de investigación sin componentes de innovaciones incrementales es más fácil. Solo hay que comparar con la investigación que realizan los países que están a la cabeza del desarrollo tecnológico. De esta forma se excluyen las empresas que producen innovaciones incrementales, cuyo volumen en los países avanzados es abrumadoramente superior al nuestro.

*1.1. Ausencia de una estrategia de funcionamiento*

España ha tenido dificultades seculares con la investigación que se han producido por la incapacidad de nuestros gobernantes para aplicar una política de investigación razonable. Para documentar esta conclusión solo hay que fijarse en las dos últimas estrategias para la investigación publicadas (Estrategia Española de Ciencia, Tecnología e innovación; EECTI a partir de aquí), las de 2013-2020 y 2021-2027. La falta de acierto de estos dos documentos para determinar la situación de la investigación pública, redactados por dos gobiernos absolutamente independientes, pone de manifiesto que las raíces del problema son muy profundas.

La falta de acierto en la descripción de lo que se tiene es un grave problema en cualquier estrategia de mejora de cualquier sistema productivo. La estrategia tiene que partir de lo que se tiene. Esto incluye a la investigación, aunque su producto, el avance del conocimiento, sea atípico en los sistemas productivos. Si no se sabe de dónde se parte, cabe esperar que tampoco se sepa a dónde hay que llegar. Eso les pasa a las dos EECTI descritas. En la primera, el sistema se describe por la inversión, el número de investigadores y el número de trabajos publicados. Muy sencillo, pero erróneo. Con respecto a los dos primeros parámetros, es evidente que ni la inversión ni el número de trabajadores sirven para definir el comportamiento de ningún sistema productivo. Para la EECTI 2013-2020, los conceptos de rendimiento, eficiencia, eficacia o productividad no existen. Su inexistencia lleva a concluir que poca estrategia puede diseñarse con tan poca base.

Con respecto al número de trabajos publicados, es evidente que este es un dato estadístico, pero absolutamente inútil para analizar el sistema. En cualquier industria, lo importante no es cuántos elementos se producen sino cuántos se venden. Sería difícil encontrar un empresario que cuente solo lo que produce y que viva satisfecho porque produce mucho, aunque no lo venda. En investigación, la pregunta pertinente es con cuánto se ha contribuido al avance global del conocimiento.

Sobre la EECTI 2021-2027, en un trabajo previo[4] ya he demostrado que contiene graves errores metodológicos que llevan a conclusiones gravemente erróneas. Por ejemplo, concluir que España está ampliamente por encima de la media en las publicaciones que están en el 10% o 1% más citas del mundo es un puro error de cálculo matemático. En resumen, en los últimos 20 o 25 años **España no ha tenido estrategia de ningún tipo para mejorar su sistema de investigación**.

*1.2. Propósito de este estudio*

Atendiendo a lo descrito, el principal propósito de este estudio es **definir la situación de la investigación en España**. Como se ha dicho antes, esta descripción es crucial como base de partida para desarrollar una certera política científica que permita incorporarnos al club de los países que basan su economía en el conocimiento. Este trabajo trata de corregir el desconocimiento de la realidad de la investigación en España y de acercar el conocimiento académico actual a la realidad de la política científica.

Con este propósito, el artículo se divide en tres partes. En la primera parte se describe cómo se mide la eficiencia de la producción científica; esta descripción se hace con cierto detalle ya que acertar en esta medida es crucial en el análisis de cualquier sistema de investigación. En la segunda parte, los métodos descritos se aplican a la evaluación de la producción científica española y a la de otros países de nuestro entorno para obtener una evaluación realista de la investigación en España. La evaluación se centra fundamentalmente en las universidades porque los datos más rigurosos son los que publica el *Leiden Ranking* y, desafortunadamente, el *Leiden Ranking* solo analiza universidades, con muy pocas excepciones. La tercera parte resume las conclusiones más importantes que se deducen de los resultados.

Por razones de espacio y de conveniencia para el lector, este artículo no trata aspectos tan importantes como la eficiencia en relación con la inversión[5] o las razones que han motivado la baja competitividad de la investigación en España. Estos son temas de máxima importancia, pero que requieren una metodología de estudio diferente de la que aquí se presenta.

**2. Las medidas de la ciencia**

*2.1. El uso del número de citas*

El éxito de un sistema científico radica en su capacidad para hacer descubrimientos que hagan avanzar el conocimiento o que puedan aplicarse a un sistema productivo. En este contexto, el éxito de un sistema científico nada tiene que ver con el número de trabajos publicados, porque un trabajo publicado puede dar origen a un premio Nobel, no aportar nada al conocimiento científico o a cualquier cosa entre estos dos extremos. Para valorar un trabajo científico, una investigación intensa y extensa ha llevado a concluir que el número de citas científicas que recibe correlaciona con su importancia. Hay que resaltar que el término matemático de correlacionar no equivale a medir y que esto produce ciertas restricciones en su uso. Desafortunadamente, en muchas evaluaciones estas restricciones se ignoran.

Un caso cotidiano ilustra este problema. En los lactantes y más allá de la lactancia, el peso de los niños correlaciona con su edad, los niños de más edad pesan más que los mas jóvenes. Sin embargo, el peso de un niño no sirve para determinar su edad, ya que podría ser un niño anormalmente grande o anormalmente pequeño. Por el contrario, si pesamos a 100 niños nacidos el mismo día, la media de su peso nos dará con bastante precisión la edad de esa población de niños (es un ejemplo al margen de errores de muestreo o diferencias entre poblaciones).

Con las citas de los trabajos la situación es muy similar. En muchos casos, el número de citas será una medida razonablemente aproximada a la importancia, pero en algunos las desviaciones pueden ser muy grandes. Siendo riguroso, la evaluación por el número de citas tiene que aplicarse solo en los niveles de agregación que conllevan un alto número de trabajos, como son los casos de la evaluación de países o instituciones.

En el caso de las universidades, el sistema de evaluación del Reino Unido[6] se realiza por expertos y se ha utilizado para validar inequívocamente los indicadores por percentiles basados en el número de citas que más abajo se describen[7,8].

*2.2. La probabilidad de hacer un descubrimiento importante*

En bibliometría, el número de indicadores propuestos es muy alto, pero pocos se han validado con las evaluaciones por expertos, como se indica en la sección anterior.

Un indicador de tipo matemático es la probabilidad de hacer un descubrimiento que signifique un notable avance del conocimiento o un avance tecnológico importante. Siguiendo la terminología de Thomas Kuhn[9], la publicaciones pueden ser *revolucionarias* o *normales*. Las primeras son las que corresponden a los descubrimientos importantes y las segundas son las aportaciones que se hacen rutinariamente. El número de las primeras es muy inferior al de las segundas, pero son estas y no las segundas las que nos interesan. Si un país o una institución solo hace *publicaciones normales*, su papel en la ciencia solo será trabajar y aportar datos para que los descubrimientos los hagan otros. Como veremos, este es el caso de España.

Como se ha dicho antes, la importancia de un trabajo correlaciona con el número de citas, de tal manera que, en países e instituciones, la probabilidad de hacer una aportación científica muy importante se puede calcular como la probabilidad de publicar un trabajo muy citado y esta probabilidad podrá calcularse si se conoce la función de distribución de las citas. Hay que notar que en las instituciones y en los países poco desarrollados, la probabilidad de publicar trabajos importantes en un año (o en otro periodo de tiempo) se puede calcular, pero los trabajos anuales muy importantes no se pueden contar porque pueden hacer falta varios años para publicar uno.

La distribución de los trabajos científicos de acuerdo con el número de citas es una función normal logarítmica como la que se presenta en la Figura 1. Hay que destacar que en estas distribuciones la cola de la derecha es larguísima. Al contrario de las distribuciones a las que estamos muy acostumbrados, que son simétricas o casi simétricas con respecto a la media, en una distribución normal logarítmica, la probabilidad de un caso que se aleja 10 o incluso 100 veces de la media es baja, pero no es despreciable. Por poner un ejemplo, si el peso medio de los niños al cumplir un año es de 9-10 kg, la probabilidad de que un niño cogido al azar pese más de 100 kg o más de 1.000 kg es cero (casi en términos matemáticos). Por el contrario, si la media de citas anuales de los trabajos publicados en una disciplina concreta es 3, la probabilidad de que un trabajo reciba más de 30 citas es baja pero notable y la probabilidad de que reciba más de 300 citas es muy baja, pero, en muchas disciplinas, cada año se publican varios de estos trabajos y son los más importantes.

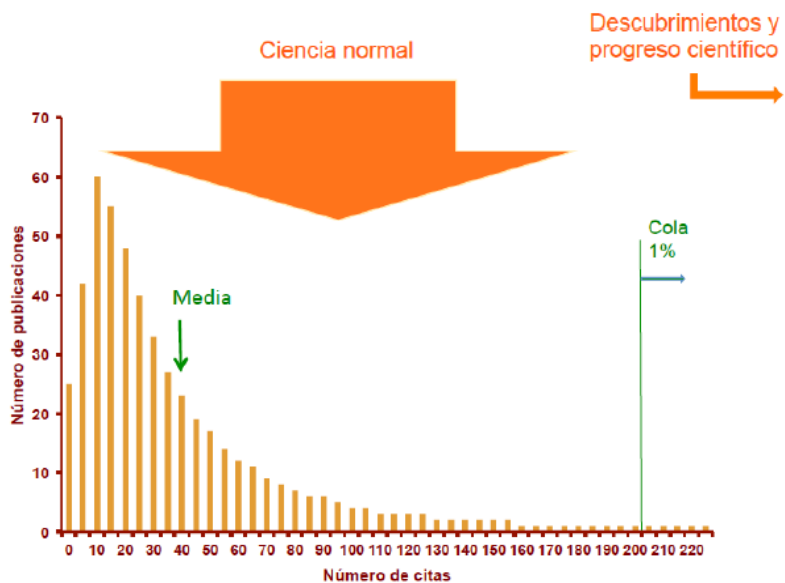

*Figura 1. Diagrama representativo de la distribución de citas de los trabajos publicados por una institución o un país. Por ejemplo, trabajos publicados un año y citas en cuatro años posteriores, sin contar el siguiente inmediato. Alrededor del 65-70% de las publicaciones quedan a la izquierda de la media. La consideración de que el 99% de las publicaciones es ciencia normal es solo una opción por el tamaño de la figura; probablemente, una proporción aún mayor sería más acertada*

Los descubrimientos aparecen en esa larga cola de las publicaciones que están a la derecha de la Figura 1, muy lejos del límite de la figura. Por eso se dice que los descubrimientos científicos son *heavy-tailed*: "*Science's heavy tail allows us to expect even greater future discoveries, even if we cannot predict when they will occur or even what fields they will occur in*"[10]. En español diríamos que la larga cola de la ciencia nos hace esperar grandes descubrimientos, aunque no sepamos predecir cuándo o en que disciplina ocurrirán.

En resumen, lo que define el rendimiento de un sistema científico son las publicaciones muy citadas, cuyo número correlaciona con el número de *descubrimientos*. Conviene tener presente que estas publicaciones no aparecen aisladas de los otros muchos trabajos poco citados: las muy citadas son parte inseparable de la producción total. Su proporción varía con la eficiencia del sistema, pero siempre son muy pocas. Un ejemplo de la variación de esta proporción la tenemos dividiendo el número de publicaciones por el número de premios Nobel recibidos en instituciones o países. En el periodo 1989-2008 en el *Massachusetts Institute of Technology* (MIT) esa ratio fue 5.544 trabajos por cada premio Nobel en ciencias, en el conjunto de EEUU la ratio fue 42.134, en Suiza fue 81.804 y en el resto de los países con premios Nobel varió entre 100.000 y 180.000[11].

Si se quiere conocer la probabilidad de publicar un trabajo muy citado, lo único que hay que hacer es contar las citas recibidas por todos los trabajos publicados en un año determinado, durante un periodo determinado después de su publicación. La distribución de los trabajos se ajusta a una función como la de la Figura 1 y el cálculo de la probabilidad es un problema matemáticamente resuelto hace muchos años[12].

No obstante, el uso del número de citas para evaluar presenta una grave complicación porque las pautas de citas dependen del área de trabajo, por razones bien establecidas que no es necesario detallar aquí; por ejemplo, los trabajos en biología molecular se citan mucho más que los de matemáticas. Incluso dentro de la misma disciplina las diferencias son grandes; por ejemplo, en tecnología, las publicaciones en semiconductores reciben muchas más citas que en metalurgia. Este problema se resuelve normalizando las citas, de tal manera que, después de normalizar, lo que se obtenga tenga el mismo significado en disciplinas diferentes.

De todos los métodos investigados para normalizar, el más robusto y el que actualmente está en uso es el de los percentiles (usado por ejemplo por la *National Science Foundation* en sus *Science and Engineering Indicators*). En cada disciplina, se crea una lista mundial de trabajos ordenados por el número de citas, empezando por el más citado. En la lista global, un percentil, por ejemplo el 1% o el 10%, incluye los trabajos de esta lista hasta llegar al límite del percentil, siempre empezando por el más citado. Para evaluar a un país o institución se cuentan los trabajos de ese país o institución que están en el grupo del mundo del percentil elegido. El número, sin más cálculos, sería un indicador dependiente del tamaño del sistema, pero si se divide por el número total de trabajos publicados nos daría un indicador independiente de tamaño que nos permitiría comparar sistemas sin tener en cuenta el tamaño. Por ejemplo, en la Universidad de Harvard el 3,3% de sus trabajos están en el 1% superior del mundo mientras que en la Universidad de Barcelona, la primera de España, solo el 1,1% de sus trabajos están en ese percentil. En la mayor parte de las universidades españolas, solo el 0,6-0,7% de sus trabajos están en el 1% superior de los trabajos del mundo. Lo que quiere decir que la mayor parte de nuestras universidades no llegan a la media mundial: los tamaños no importan en esta comparación.

Con los datos de los percentiles también se pueden calcular probabilidades. En este caso, en lugar de calcular la probabilidad de que un trabajo tenga más de un número determinado de citas, se calcula la probabilidad de que esté en un percentil determinado. Por ejemplo, que esté en el 1% o el 0.1% superior de la lista de las publicaciones mundiales. El procedimiento matemático está publicado[13] y no hace falta detallarlo aquí.

Haciendo uso de procedimientos analíticos de diversa índole, investigadores del Max Planck en Alemania[14] han calculado que solo el 0,02% de las publicaciones alcanzan el nivel de publicaciones que han hecho aportaciones importantes al progreso científico. El nivel de premio Nobel es más exigente, el 0,001%, aproximadamente. Con esa proporción, si actualmente se publican tres millones de trabajos anuales (*Web of Science*), solo entre 30 y 90 publicaciones tienen el nivel de premio Nobel. Estas proporciones son sobre la producción global, en áreas específicas las proporciones varían.

*2.2. Los listados de Ioannidis*

En cualquier disciplina, los avances académicos tardan en llegar a la vida real, pero en política científica parece que nuca llegan. Las EECTI españolas son un ejemplo de cómo los responsables políticos ignoran a los investigadores académicos y evalúan la investigación utilizando aproximaciones e indicadores erróneos, alejados del estado del conocimiento en la materia. Hacer esto conduce inexorablemente a la confusión y a políticas equivocadas. Algo parecido también pasa en otros países europeos[15], pero en menor medida.

Una razón que podría explicar el uso de indicadores inadecuados por los responsables políticos es el rechazo a las matemáticas, aunque sean sencillas. Por el contrario, contar el número de algo que tenga un significado de éxito científico muy claro—podría ser el número de premios Nobel en instituciones muy acreditadas—es más atractivo que calcular probabilidades. Un indicador de este tipo es el número de investigadores muy citados que tiene una institución o un país—las matemáticas solo las tiene que usar el que hace los conteos— y esos conteos existen y se pueden utilizar.

John Ioannidis (Universidad de Stanford) y colaboradores estudiaron la producción científica de 6,9 millones de investigadores creando un índice compuesto basado en el número de citas de sus trabajos. En 2019 seleccionaron los 100.000 investigadores con un mayor índice y en 2020 lo completaron hasta alcanzar el 2% en cada disciplina[16]. Ya se ha comentado que a niveles bajos de agregación de publicaciones (p. e., autores) los índices bibliométricos pueden fallar, pero si se agregan para una institución o un país los índices bibliométricos son robustos.

En resumen, contando los investigadores que aparecen en las listas de John Ioannidis y colaboradores[16] (*investigadores Ioannidis*) que hay en un país o una institución tendremos una medida muy razonable del nivel de su investigación. Es intuitivo que las instituciones con mayor número de investigadores con éxito serán las que tienen más éxito y el uso de este número para estimar el éxito de una institución tendría que ser fácilmente aceptado por los responsables de la política científica.

El número de *investigadores Ioannidis* correlaciona muy bien con los datos del *Leiden Ranking*; y para las universidades españolas, contar el número de *investigadores Ioannidis* tiene algunas ventajas sobre los cálculos de probabilidad realizados a partir de los datos del *Leiden Ranking*. La razón es que en muchas universidades españolas el número de citas a las publicaciones es bajo y los cálculos de probabilidad partiendo de los datos del *Leiden Ranking* son menos exactos que los cálculos de Ioannidis, que se basan en periodos más largos.

Cuando se comparan estos dos tipos de datos hay que tener presente que la exigencia para los *investigadores Ioannidis*, aproximadamente el 2% más acreditado de la población, es muy inferior a la exigencia para las publicaciones que están en el 0,02% más acreditado.

**3. El nivel de la investigación en España**

El *Leiden Ranking* contiene numerosos datos de 41 universidades españolas que satisfacen unos requerimientos mínimos en investigación. Como ya se ha dicho, con estos datos se puede calcular la probabilidad de que una publicación al azar de esas universidades alcance el nivel del 0,02% más citado, lo que corresponde a un trabajo científico muy importante. Este es un indicador que permite comparar universidades, independientemente del tamaño que tengan, pero que no da idea del número de *descubrimientos* que puede hacer la universidad, porque este último es un indicador dependiente del número de trabajos que se publican. Multiplicando el número total de publicaciones por la probabilidad arriba descrita obtenemos el número esperable de artículos muy importantes.

**Tabla1. Actividad investigadora de las universidades españolas que aparecen en el *Leiden Ranking*: probabilidad de publicar un trabajo que se encuentre en el 0,02% más citado de los trabajos publicados en el mundo, número de publicaciones, número de publicaciones en el percentil 0,02 (avances) y número de investigadores Ioannidis**

| Universidad | Probabilidad 0,02 (x1000) | Número de Publicaciones | Número de avances | Número Ioannidis |
|---|---|---|---|---|
| Barcelona | 028 | 6.015 | 1,71 | 140 |
| Complutense de Madrid | 0,08 | 5.154 | 0,42 | 78 |
| Autónoma de Barcelona | 0,17 | 5.102 | 0,86 | 70 |
| Politécnica de Cataluña | 0,11 | 3.401 | 0,37 | 68 |
| Valencia | 0,12 | 4.668 | 0,56 | 64 |
| Zaragoza | 0,06 | 3.186 | 0,20 | 55 |
| País Vasco | 0,15 | 3.945 | 0,60 | 51 |
| Granada | 0,16 | 4.380 | 0,71 | 49 |
| Autónoma de Madrid | 0,24 | 3.885 | 0,92 | 47 |
| Politécnica de Valencia | 0,13 | 3.776 | 0,50 | 46 |
| Sevilla | 0,10 | 4.124 | 0,41 | 46 |
| Santiago de Compostela | 0,10 | 2.826 | 0,28 | 39 |
| Pompeu Fabra | 0,50 | 1.355 | 0,68 | 34 |
| Oviedo | 0,07 | 2.491 | 0,17 | 29 |
| Alicante | 0,04 | 1.710 | 0,06 | 28 |
| Murcia | 0,09 | 2.115 | 0,19 | 27 |
| Politécnica de Madrid | 0,10 | 3.123 | 0,33 | 25 |
| Extremadura | 0,05 | 1.349 | 0,07 | 23 |
| Salamanca | 0,08 | 1.614 | 0,13 | 21 |
| Vigo | 0,08 | 1.841 | 0,14 | 20 |
| Alcalá | 0,08 | 1.239 | 0,09 | 19 |
| Córdoba | 0,09 | 1.650 | 0,15 | 19 |
| Gerona | 0,37 | 1.149 | 0,42 | 18 |
| Málaga | 0,10 | 1.925 | 0,19 | 18 |
| Islas Baleares | 0,32 | 1.137 | 0,37 | 18 |
| Castilla-La Mancha | 0,06 | 2.055 | 0,12 | 17 |
| Valladolid | 0,04 | 1.574 | 0,07 | 16 |
| Rovira i Virgili | 0,21 | 1.608 | 0,33 | 14 |
| Lérida | 0,19 | 838 | 0,16 | 14 |
| Carlos III de Madrid | 0,12 | 1.560 | 0,18 | 13 |
| Cantabria | 0,13 | 1.203 | 0,16 | 13 |
| Navarra | 0,34 | 1.255 | 0,43 | 13 |
| Almería | 0,05 | 888 | 0,04 | 10 |
| Rey Juan Carlos | 0,16 | 1.013 | 0,16 | 9 |
| Miguel Hernández | 0,09 | 1.068 | 0,10 | 8 |
| Jaume I | 0,10 | 1.216 | 0,12 | 8 |
| La Laguna | 0,06 | 1.179 | 0,07 | 7 |
| Las Palmas de Gran Canaria | 0,03 | 929 | 0,02 | 7 |
| La Coruña | 0,09 | 1.194 | 0,10 | 6 |
| Cádiz | 0,05 | 1.095 | 0,06 | 5 |
| Jaén | 0,04 | 996 | 0,04 | 3 |
| **Total** | | | **12,72** | **1215** |

Los valores de las probabilidaes se han calculado a partir de los datos del *Leiden Ranking* 2020 (conteo fraccionario; periodo 2015-2018) con el procedimiento descrito en Rodríguez-Navarro A, Brito R (2019). Probability and expected frequency of breakthroughs: basis and use of a robust method of research assessment, *Scientometrics* 119: 213-235

Los números de investigadores Ioannidis se han obtenido de la publicación Ioannidis JPA, Boyack KW, Baas J (2020), Updated science-wide author databases of standardized citation indicators, *PLoS Biol* 18(10):e3000918

Con referencia al número de *investigadores Ioannidis*, prácticamente todas la universidades recogidas en el *Leiden Ranking* tienen investigadores que están en los listados de Ioannidis y solo hay que contarlos para tener un indicador adicional al calculado de los datos del *Leiden Ranking*, como ya se ha explicado.

La Tabla 1 recoge todos estos datos para las 41 universidades españolas que aparecen en el *Leiden Ranking*: la probabilidad de que una publicación esté en el 0,02% más citado, el número de trabajos publicados, el número esperable de trabajos que alcanzarían el percentil 0.02 y el número de *investigadores Ioannidis*.

Como esos indicadores son difíciles de interpretar en términos absolutos, la Tabla 2 recoge los mismos indicadores de la Tabla 1 para universidades fuera de España; en general, las dos universidades más conocidas de cada país. En primer lugar están dos países, Australia y Canadá, cuyo PIB son similares al de España. Luego se incluyen siete países europeos con diversos niveles en investigación[15]. También se incluyen dos universidades de EEUU de gran prestigio, la Universidad de Harvard y el MIT. Finalmente, considerando la popularidad del estado de California en tecnologías avanzadas, se incluyen cuatro campus de la Universidad de California: Berkeley, San Francisco, Los Ángeles y San Diego.

Tabla 2. Actividad investigadora de algunas universidades no españolas: probabilidad de publicar un trabajo que se encuentre en el 0,02% más citado de los trabajos publicados en el mundo, número de publicaciones, número esperable de publicaciones en el percentil 0,02 (avances) y número de investigadores Ioannidis

| Universidad | País | Probabilidad 0,02 (x1,000) | Número de publicaciones | Número de avances | Número Ioannidis |
|---|---|---|---|---|---|
| University of Queensland | Australia | 0,59 | 12.316 | 7.23 | 393 |
| University of Sydney | Australia | 0,56 | 12.604 | 7,05 | 421 |
| University of Toronto | Canada | 0,69 | 22.995 | 15,7 | 933 |
| University of British Columbia | Canada | 0,59 | 12.988 | 7,65 | 687 |
| Ludwig-Maximilians-Universität München | Alemania | 0,57 | 7.409 | 4,19 | 229 |
| University of Freiburg | Alemania | 0,32 | 4.923 | 1,59 | 193 |
| Karolinska Institutet | Suecia | 0,51 | 8.324 | 4,28 | 287 |
| Lund University | Suecia | 0,30 | 8.181 | 2,44 | 258 |
| Sorbonne University | Francia | 0,43 | 8.767 | 3,80 | 254 |
| Université Paris-Saclay | Francia | 0,46 | 8.235 | 3,76 | 214 |
| University of Padova | Italia | 0,24 | 7.678 | 1,83 | 214 |
| University of Bologna | Italia | 0,21 | 7.271 | 1,51 | 190 |
| Utrecht University | Países Bajos | 1,04 | 9.391 | 9,81 | 196 |
| University of Amsterdam | Países Bajos | 0,95 | 9.081 | 8,66 | 209 |
| Eidgenössische Technische Hochschule (ETH) Zürich | Suiza | 1,88 | 9.342 | 17,6 | 406 |
| Ecole Polytechnique Fédérale de Lausanne | Suiza | 1,85 | 5.506 | 10,1 | 233 |
| University of Cambridge | Reino Unido | 1,90 | 13.485 | 25,6 | 605 |
| University of Oxford | Reino Unido | 2,00 | 15.353 | 30,7 | 801 |
| Harvard University | EEUU | 3,42 | 33.722 | 115,3 | 1.510 |
| Massachusetts Institute of Technology (MIT) | EEUU | 5,48 | 10.563 | 57,9 | 619 |
| University of California, Berkeley | EEUU | 3,31 | 10.671 | 35,2 | 734 |
| University of California, San Francisco | EEUU | 2,21 | 9.994 | 22,1 | 686 |
| University of California, Los Angeles | EEUU | 1,29 | 13.645 | 17,5 | 634 |
| University of California, San Diego | EEUU | 1,71 | 12.135 | 20,7 | 474 |

Los valores de las probabilidades se han calculado a partir de los datos del *Leiden Ranking* 2020 (conteo fraccionario; period 2015-2018) con el procedimiento descrito en Rodríguez-Navarro A, Brito R (2019). Probability and expected frequency of breakthroughs: basis and use of a robust method of research assessment, *Scientometrics* 119: 213-235,

Los números de investigadores Ioannidis se han obtenido de la publicación Ioannidis JPA, Boyack KW, Baas J (2020), Updated science-wide author databases of standardized citation indicators, *PLoS Biol* 18(10):e3000918,

Atendiendo a las publicaciones muy importantes, está claro que las universidades españolas quedan bastante por detrás de las universidades de los países europeos menos competitivos y muy por detrás de los países europeos más competitivos: Países Bajos, Suiza y Reino Unido. Por ejemplo, fijándonos en los datos correspondientes al periodo 2015-2018, el conjunto de todas las universidades españolas habrá producido 13 avances científicos importantes, que es **la mitad de lo que corresponde a la Universidad de Cambridge, cuatro veces menos que el MIT y nueve veces menos que la Universidad de Harvard**.

Atendiendo al número de *investigadores Ioannidis*, los resultados son similares. En el conjunto de las universidades españolas hay 1.215, que es solo el doble de los que se cuentan en la Universidad de Cambridge y menos que los que hay en la Universidad de Harvard. Ya se ha explicado que el nivel de los *investigadores Ioannidis* corresponde al 2% superior y que las diferencias son menores que las que se aprecian en los trabajos esperables en el percentil 0,02% superior, que corresponde a los descubrimientos importantes.

El número de *investigadores Ioannidis* en las universidades españolas es aproximadamente la mitad del total de España, donde hay 2.291. En el CSIC hay 393 *investigadores Ioannidis*, que son bastantes menos de los que hay en muchas universidades europeas (Tabla 2) que, además, tienen una docencia muy activa.

Una idea clara de la situación de España en su conjunto se puede obtener contando el número de *investigadores Ioannidis* y dividiendo por el número de habitantes (Tabla 3). En España tenemos 49 investigadores por millón de habitantes mientras que los países avanzados tiene alrededor de 200. Estamos a la cola de Europa, mejor que Portugal, pero peor que Grecia.

**Tabla 3, Número de investigadores Ioannidis en España y en otros países seleccionados y ratio de investigadores Ioannidis por millón de habitantes**

| País | Número Ioannidis | Habitantes (millones) | Ratio |
|---|---|---|---|
| Suiza | 2.545 | 8,6 | 296 |
| Dinamarca | 1.494 | 5,8 | 258 |
| Suecia | 2.545 | 10,3 | 247 |
| Gran Bretaña | 15.002 | 64 | 234 |
| Australia | 5.440 | 25 | 218 |
| EEUU | 68.015 | 328,2 | 207 |
| Países Bajos | 3.352 | 17 | 197 |
| Canadá | 7.224 | 37,6 | 192 |
| Nueva Zelanada | 802 | 5 | 160 |
| Bélgica | 1.411 | 11,4 | 124 |
| Austria | 961 | 8,9 | 108 |
| Alemania | 8.791 | 83 | 106 |
| Francia | 5.011 | 67 | 75 |
| Italia | 4.006 | 60 | 67 |
| Grecia | 647 | 10,7 | 60 |
| España | 2.291 | 47 | 49 |
| Portugal | 384 | 10,3 | 37 |

Los números de investigadores Ioannidis se han obtenido de la publicación Ioannidis JPA, Boyack KW, Baas J (2020), Updated science-wide author databases of standardized citation indicators, *PLoS Biol* 18(10):e3000918,

**4. Evolución de la investigación en las universidades españolas**

El crecimiento del número de publicaciones en España entre 1990 y 2005 fue espectacular[17], pero no hay constancia de ningún tipo de mejora en la probabilidad de producir avances. En los últimos años, a pesar de los recortes de financiación, la producción ha seguido aumentando, pero la calidad no ha mejorado.

**Tabla 4. Evolución de la actividad investigadora de las universidades españolas: probabilidad de publicar un trabajo que se encuentre en el 0,02% más citado de los trabajos publicados en el mundo, número de publicaciones y número esperable de trabajos en el 0,02% más citado (avances).**

| Universidad | Probabilidad 0,02% (x1000) | | Número de publicaciones | | Número de avances | |
|---|---|---|---|---|---|---|
| | 2006-2009 | 2015-2018 | 2006-2009 | 2015-2018 | 2006-2009 | 2015-2018 |
| Barcelona | 0,20 | 0,28 | 5148 | 6015 | 1,04 | 1,71 |
| Complutense de Madrid | 0,09 | 0,08 | 4371 | 5154 | 0,39 | 0,42 |
| Autónoma de Barcelona | 0,15 | 0,17 | 3848 | 5102 | 0,57 | 0,86 |
| Politécnica de Cataluña | 0,13 | 0,11 | 2528 | 3401 | 0,33 | 0,37 |
| Valencia | 0,11 | 0,12 | 3482 | 4668 | 0,37 | 0,56 |
| Zaragoza | 0,16 | 0,06 | 2506 | 3186 | 0,39 | 0,20 |
| País Vasco | 0,11 | 0,15 | 2438 | 3945 | 0,26 | 0,60 |
| Granada | 0,10 | 0,16 | 2900 | 4380 | 0,28 | 0,71 |
| Autónoma de Madrid | 0,21 | 0,24 | 3314 | 3885 | 0,68 | 0,92 |
| Politécnica de Valencia | 0,17 | 0,13 | 2441 | 3776 | 0,42 | 0,50 |
| Sevilla | 0,12 | 0,10 | 2611 | 4124 | 0,32 | 0,41 |
| Santiago de Compostela | 0,15 | 0,10 | 2670 | 2826 | 0,40 | 0,28 |
| Pompeu Fabra | 0,48 | 0,50 | 668 | 1355 | 0,32 | 0,68 |
| Oviedo | 0,06 | 0,07 | 1967 | 2491 | 0,12 | 0,17 |
| Alicante | 0,33 | 0,04 | 1225 | 1710 | 0,40 | 0,06 |
| Murcia | 0,05 | 0,09 | 1737 | 2115 | 0,08 | 0,19 |
| Politécnica de Madrid | 0,04 | 0,10 | 1801 | 3123 | 0,07 | 0,33 |
| Extremadura | 0,06 | 0,05 | 1115 | 1349 | 0,07 | 0,07 |
| Salamanca | 0,04 | 0,08 | 1471 | 1614 | 0,05 | 0,13 |
| Vigo | 0,14 | 0,08 | 1516 | 1841 | 0,22 | 0,14 |
| Alcalá | 0,07 | 0,08 | 916 | 1239 | 0,06 | 0,09 |
| Córdoba | 0,13 | 0,09 | 1202 | 1650 | 0,16 | 0,15 |
| Gerona | 0,27 | 0,37 | 688 | 1149 | 0,18 | 0,42 |
| Málaga | 0,04 | 0,10 | 1187 | 1925 | 0,05 | 0,19 |
| Islas Baleares | 0,33 | 0,32 | 794 | 1137 | 0,26 | 0,37 |
| Castilla-La Mancha | 0,12 | 0,06 | 1393 | 2055 | 0,17 | 0,12 |
| Valladolid | 0,06 | 0,04 | 1200 | 1574 | 0,07 | 0,07 |
| Rovira i Virgili | 0,26 | 0,21 | 1163 | 1608 | 0,31 | 0,33 |
| Lérida | 0,10 | 0,19 | 516 | 838 | 0,05 | 0,16 |
| Carlos III de Madrid | 0,14 | 0,12 | 909 | 1560 | 0,13 | 0,18 |
| Cantabria | 0,07 | 0,13 | 962 | 1203 | 0,07 | 0,16 |
| Navarra | 0,05 | 0,34 | 1201 | 1255 | 0,06 | 0,43 |
| Almería | 0,12 | 0,05 | 562 | 888 | 0,07 | 0,04 |
| Rey Juan Carlos | 0,16 | 0,16 | 644 | 1013 | 0,10 | 0,16 |
| Miguel Hernández | 0,09 | 0,09 | 722 | 1068 | 0,06 | 0,10 |
| Jaume I | 0,15 | 0,10 | 636 | 1216 | 0,09 | 0,12 |
| La Laguna | 0,02 | 0,06 | 1042 | 1179 | 0,02 | 0,07 |
| Las Palmas de Gran Canaria | 0,03 | 0,03 | 521 | 929 | 0,01 | 0,02 |
| La Coruña | 0,03 | 0,09 | 714 | 1194 | 0,02 | 0,10 |
| Cádiz | 0,07 | 0,05 | 633 | 1095 | 0,04 | 0,06 |
| Jaén | 0,03 | 0,04 | 611 | 996 | 0,02 | 0,04 |
| **Total** | | | 67972 | 92830 | 8,80 | 12,72 |
| **Media** | 0,13 | 0,14 | | | | |

Los valores de las probabilidades se han calculado a partir de los datos del *Leiden Ranking* 2020 (conteos fraccionarios) con el procedimiento descrito en Rodríguez-Navarro A, Brito R (2019). Probability and expected frequency of breakthroughs: basis and use of a robust method of research assessment. *Scientometrics* 119: 213-235.

En la Tabla 4 se han incluido dos periodos de cálculos del *Leiden Ranking*: 2006-2009, el primero, y 2015-2018, el último. Atendiendo al número de publicaciones, todas las universidades están en ascenso. Pero si atendemos a la probabilidad de publicar un trabajo en el 0,02% más citado, el cambio es muy variable. Atendiendo al conjunto, el progreso ha sido muy pequeño, de $0,13 \cdot 10^{-3}$ a $0,14 \cdot 10^{-3}$. Pero, más importante, el progreso desaparece si se eliminan tres universidades: Barcelona, Autónoma de Barcelona y Autónoma de Madrid (eliminándolas, la probabilidad es $0,12 \cdot 10^{-3}$ en los dos periodos).

**5. Las universidades tecnológicas**

Muchas tecnologías requieren una investigación muy activa. En algunas tecnologías (por ejemplo automoción) las innovaciones son fundamentalmente incrementales, pero en muchas tecnologías modernas las innovaciones son revolucionarias. Por ejemplo, en el caso arriba mencionado de las baterías eléctricas, la investigación ha permitido avances espectaculares tanto en la capacidad como en la rapidez de la carga y esto ha tenido lugar a partir de innovaciones revolucionarias (los premios Nobel de Química en 2019 son un ejemplo). Estas consideraciones llevan de nuevo a preguntarse cuál es la situación de la investigación en tecnología en España.

Un problema que aparece cuando se utilizan análisis para toda la ciencia en general, como en las Tablas 1 y 2, es la falta de especificidad en áreas tecnológicas. Las universidades no cubren todas las áreas del conocimiento con igual intensidad y la normalización matemática de los índices no alcanza a normalizar la singularidad de las áreas de conocimiento con respecto a las citas (por ejemplo, en algunas áreas se publica más en libros que en artículos de revista, pero solo en estos últimos es fácil seguir las citas). Una forma de centrarse más en la tecnología y homogeneizar el espectro de áreas cubiertas por las universidades es centrarse en las universidades que se declaran técnicas o politécnicas. Además, en estas universidades no existen muchas de las áreas que se evalúan mal por el número de citas.

En la Tabla 1 aparecen tres universidades de este tipo en España: Politécnica de Cataluña, Politécnica de Valencia y Politécnica de Madrid. Si nos fijamos en el número esperable de artículos muy importantes ("avances" en la Tabla 1), estas tres universidades representan el nivel medio en España. En todo el mundo, también en Europa, hay universidades que se declaran técnicas o politécnicas y que están más enfocadas a la docencia que a la investigación. Este no es el caso de las tres universidades politécnicas españolas mencionadas y lo lógico es compararlas con universidades técnicas y politécnicas que se declaren instituciones de investigación en otros países.

**Tabla 5. Actividad investigadora en universidades técnicas y politécnicas: número de publicaciones, probabilidad de publicar un trabajo que se encuentre en el 0,02% más citado de los trabajos publicados en el mundo, número de trabajos en este percentil (avances) y número de investigadores Ioannidis**

| Universidad | Número de publicaciones | Prob. 0.02% (x 1000) | Número de avances | Número Ioannidis |
|---|---|---|---|---|
| Massachusetts Institute of Technology (MIT) | 10.573 | 5,49 | 57,99 | 619 |
| Eidgenössische Technische Hochschule (ETH) Zürich | 9.342 | 1,88 | 17,60 | 406 |
| Ecole Polytechnique Fédérale de Lausanne | 5.506 | 1,85 | 10,16 | 233 |
| Technical University of Denmark | 5.860 | 0,56 | 3,29 | 246 |
| Technical University of Munich | 8.142 | 0,43 | 3,48 | 174 |
| Universidad Politécnica de Cataluña | 3.401 | 0,11 | 0,37 | 68 |
| Universidad Politécnica de Valencia | 3.776 | 0,13 | 0,50 | 46 |
| Universidad Politécnica de Madrid | 3.123 | 0,11 | 0,33 | 25 |

Los valores de las probabilidades se han calculado a partir de los datos del *Leiden Ranking* 2020 (conteo fraccionario; period 2015-2018) con el procedimiento descrito en Rodríguez-Navarro A, Brito R (2019). Probability and expected frequency of breakthroughs: basis and use of a robust method of research assessment. *Scientometrics* 119: 213-235.

Los numerous de los investigadores Ioannidis se han obtenido de la publicación Ioannidis JPA, Boyack KW, Baas J (2020). Updated science-wide author databases of standardized citation indicators. *PLoS Biol* 18(10):e3000918.

La Tabla 5 compara las tres universidades españolas con cuatro universidades europeas de referencia y con el MIT. Los datos son elocuentes, las universidades españolas tendrían que trabajar cien veces más tiempo que el MIT para producir el mismo avance del conocimiento. Incluso en el caso de universidades de menor éxito científico, como son las Universidades Técnicas de Múnich y Dinamarca, las diferencias son de uno a diez, aproximadamente. Las diferencias en el número de *investigadores Ioannidis* no son tan acusadas porque este es un indicador menos exigente. Si tomamos un caso en el que los indicadores coinciden: *Technical University of Denmark* y Universidad Politécnica de Madrid, llegamos a la conclusión de que la inversión en la segunda tendría que ser diez veces superior para producir los mismos avances del conocimiento que la primera.

Estos resultados indican el bajo nivel de la investigación tecnológica en España, en concordancia con lo ya descrito.

**6. Resumen y conclusiones**

1. La Tablas 1 y 2 describen con claridad una baja competitividad de la investigación en las universidades españolas; la Tabla 3 describe la situación en el nivel global del país y la Tabla 5 la situación en las universidades tecnológicas. Del análisis de los datos presentados en esas tablas solo cabe una conclusión: **el éxito de la investigación en España es muy bajo y no existen paliativos para esta conclusión**. Lamentablemente, la producción científica española, a menudo formalmente rigurosa, es ciencia normal que solo sirve para que otros hagan los avances científicos.

2. Los datos que llevan a la conclusión anterior se han obtenido utilizando fuentes radicalmente diferentes, porque los métodos de conteo y cálculo del *Leiden Ranking* y los de John Ioannidis y colaboradores son radicalmente diferentes. Incluso las bases de datos son diferentes: el *Leiden Ranking* utiliza la *Web of Science* y John Ioannidis utiliza *Scopus*. Así y todo, los resultados son totalmente coherentes, teniendo en cuenta que el éxito en las listas de Ioannidis está en el 2% superior y los cálculos de probabilidades basados en el *Leiden Ranking* se sitúan en el 0,02% superior. Hay que destacar que el método basado en los datos del *Leiden Ranking* están validados con las revisiones por expertos[6,7], lo que quiere decir que lo que se deduce las Tablas 1 y 2 es lo que se obtendría en una evaluación por expertos.

3. Es un hecho demostrado que países como Alemania, Francia e Italia tienen una investigación científica poco competitiva y que la Unión Europea ha fracasado en sus políticas científicas[15]. Pero el caso de España es mucho peor. Aunque en su conjunto la investigación en Francia y Alemania está por debajo de las expectativas, su larga historia de **potencias** científicas les permite mantener núcleos muy **potentes** que consiguen premios Nobel con cierta frecuencia. Además, su industria es muy **potente**, lo que conlleva una gran volumen de innovación incremental. Conviene tener presente que mucho del progreso que observamos en la tecnología que usamos en la vida cotidiana responde a innovaciones incrementales.

4. En España no tenemos una larga historia de éxitos científicos ni una industria potente que nos asegure progreso tecnológico con independencia de la investigación pública. Por eso, España necesita optimizar un sistema de investigación empezando desde valores muy bajos, como están haciendo China, Taiwán, Singapur y otros países asiáticos. Más arriba he descrito que "fijándonos en los datos correspondientes al periodo 2015-2018, el conjunto de todas las universidades españolas habrá producido 13 avances científicos importantes, **que es la mitad de lo que corresponde a la Universidad de Cambridge, cuatro veces menos que el MIT y nueve veces menos que la Universidad de Harvard**".

5. En estas circunstancias, si quisiéramos igualarnos con Canadá, Australia o Países Bajos en descubrimientos por millón de habitantes (Tabla 3), tendríamos que invertir alrededor del 10% del PIB en investigación. Algo que es obviamente imposible. **Sin mejorar la eficiencia de nuestra investigación, España nunca se incorporará al club de los países tecnológicamente avanzados. En términos económicos probablemente se llegaría a concluir que, lamentablemente, la mejor solución para España sería externalizar la investigación**, algo absolutamente impensable.